# Turing's Children: Representation of Sexual Minorities in STEM

Dario Sansone[1]

Christopher S. Carpenter[2]

This version: May 2020

**Abstract**

We provide the first nationally representative estimates of sexual minority representation in STEM fields by studying 142,641 men and women in same-sex couples from the 2009-2018 American Community Surveys. These data indicate that men in same-sex couples are 12 percentage points less likely to have completed a bachelor's degree in a STEM field compared to men in different-sex couples; there is no gap observed for women in same-sex couples compared to women in different-sex couples. The STEM gap between men in same-sex and different-sex couples is larger than the STEM gap between white and black men but is smaller than the gender STEM gap. We also document a gap in STEM occupations between men in same-sex and different-sex couples, and we replicate this finding using independently drawn data from the 2013-2018 National Health Interview Surveys. These differences persist after controlling for demographic characteristics, location, and fertility. Our findings further the call for interventions designed at increasing representation of sexual minorities in STEM.

[1] Vanderbilt University and University of Exeter. E-mail: dario.sansone@vanderbilt.edu
[2] Vanderbilt University, NBER and IZA. E-mail: christopher.s.carpenter@vanderbilt.edu

## 1. Introduction

In this paper, we provide the first nationally representative estimates of the representation of sexual minorities in Science, Technology, Engineering, and Mathematics (STEM).[3] By doing so, we start to address the dire need for statistics on sexual and gender minorities in STEM emphasized in the letter sent to the National Science Foundation (NSF) by 251 scientists, engineers, legal and public policy scholars, as well as 17 scientific organizations (1). Prior research has documented the presence of substantial gaps in STEM degree completion and occupational attainment in STEM fields associated with gender and race/ethnicity (2, 3), but to our knowledge there have been only a handful of studies (based on non-random samples) on STEM representation for sexual minorities.

Despite improvements in the legislative and institutional background for LGBTQ people, such as the legalization of same-sex marriage in numerous countries in the last twenty years, the workplace environment for LGBTQ scientists is still far from welcoming. In 25 U.S. states, it is still legal to discriminate against applicants and employees in hiring, firing, wages, and promotions based on their sexual orientation or gender identity (4). While the NSF tracks the participation rates of women, racial and ethnic minorities, and persons with disabilities in science and engineering (2), it does not routinely collect statistics on LGBTQ people.[4] Other federal agencies, such as the National Institutes of Health, have historically funded only a very small fraction of LGBTQ-related projects (5). Researchers have documented under-representation and worse workplace experiences for LGBT employees in STEM-related federal agencies (6). In addition, several studies and reports have documented the academic and social isolation, as well as the heterosexist and uncomfortable workplace climate faced by LGBTQ STEM professionals (7–12), in addition to explicit anti-LGBTQ harassment (13, 14). Similar experiences have been documented in the medical profession (15, 16).

Previous studies on sexual minorities in STEM are rare and have relied on small or convenience-based samples (8, 10, 17, 18). One prior study (19) used data from a 2015 survey of undergraduates that contained 147 self-identified gay men: it found that, conditional on reporting a STEM major aspiration upon college entry, gay men were 14 percentage points less likely than straight men to persist in STEM majors by the fourth year of college (even if they were more likely to have worked in a lab).

Our study builds on this prior work in two critical ways. First, we use samples of sexual minorities that are two orders of magnitude larger than previous studies (19). Specifically, we draw on data from the 2009-2018 American Community Surveys (ACS) which identify over 142,000 individuals in same-sex cohabiting romantic relationships. A large body of research in social science and demography confirms that the vast majority of same-sex couples in the ACS are gay men and lesbians (20). The ACS also contain information on the undergraduate major(s) for individuals who obtained bachelor's degrees, as well as detailed information on current occupation.[5] Both of these sources of information allow us to identify individuals in STEM fields. The large sample sizes of the ACS provide increased confidence in the magnitude and precision of estimated gay/straight gaps and permit sub-group analyses (e.g., by race).

---

[3] Throughout, we use the term 'sexual minorities' to refer to individuals who describe themselves as lesbian, gay, bisexual, queer, or 'something else'. We also refer to this population as 'LGBQ' for lesbian, gay, bisexual, and queer. Due to data limitations, we are unable to study transgender individuals (i.e., people whose gender identity and/or expression does not match their sex assigned at birth). Some studies in the literature (6) use data that include both sexual minorities and gender minorities; in those cases, we refer to the LGBTQ population (i.e., including transgender individuals).

[4] However, the NSF is currently considering adding questions about sexual orientation and gender identity in its workforce surveys (29).

Second, we complement the ACS with information from the 2013-2018 National Health Interview Surveys (NHIS) which also contain detailed information on occupation as well as direct individual-level questions about sexual orientation. This allows us to examine whether sexual minority representation in STEM fields differs between lesbian and bisexual women, for example. Sample sizes in the NHIS are smaller than in the ACS, though they are still an order of magnitude larger than prior work (4,763 self-identified sexual minorities in the 2013-2018 NHIS).

Using ACS data, we show that men in same-sex couples are significantly less likely than men in different-sex couples to receive a bachelor's degree in a STEM field, or to work in a STEM occupation. The gap in STEM degrees between men in same-sex and different-sex couples (12 percentage points) is larger in size than the STEM gap between white and black men (4 percentage points) but is smaller than the gender STEM gap (21 percentage points).[6] Consistent with the ACS results, we find that gay male respondents in the NHIS are less likely than heterosexual men to work in STEM fields. While we find comparable gaps between heterosexual men and bisexual men or men who identify with another sexual orientation category, these differences are not statistically significant. In both datasets, we find that all women are underrepresented: women are always less likely to study or work in a STEM field, irrespective of their sexual orientation.

As with prior evidence on STEM gaps associated with gender and race, our findings on LGBQ-related STEM gaps are important because addressing these gaps could increase efficiency by improving group decision-making, company performance, and the quality and variety of scientific work (21). In addition, increasing the number of LGBQ people entering into STEM fields could help to alleviate the chronic shortage of workers in these fields (22, 23).

---

[5] We also note that the ACS measures degree completion, while (19) studies persistence in STEM by the fourth year of college but does not directly observe degree completion.

[6] Because we can only identify sexual minorities in couples in the ACS data, we have compared the gap between individuals in same-sex couples and individuals in different-sex couples to other couples-based gaps (i.e., black men in couples versus white men in couples and men in couples versus women in couples). The race and gender gaps in STEM degrees are very similar if we consider all adults (i.e., if we do not restrict attention to individuals in couples), and the qualitative ordering remains true: the black/white gap in STEM degrees among men is smaller than the gap in STEM degrees between men in same-sex couples and men in different-sex couples, which itself is smaller than the gender gap in STEM degrees.

## 2. Data and methodology

### 2.1 The American Community Survey (ACS)

The main dataset used in our analysis is the ACS: a nationally representative and repeated cross-sectional survey conducted by the U.S. Census Bureau. It contains demographic, economic, social, and housing information on 1 percent of the U.S. population (or approximately 3 million people each year). Such large sample sizes facilitate studies on relatively small subpopulations, such as individuals in same-sex couples and/or working in STEM occupations, or even heterogeneity analyses among these subgroups (e.g., by sex or race within same-sex couples). These data are publicly available through IPUMS-USA at the University of Minnesota (24).

The ACS does not directly ask individuals about their sexual orientation. To identify sexual minorities, we follow a large body of prior research that uses intrahousehold relationships to identify individuals in same-sex couples (25). Specifically, the ACS identifies a primary reference person, defined as "the person living or staying here in whose name this house or apartment is owned, being bought, or rented". For each individual in the household, the ACS also collects information on sex and the individual's relationship to the primary reference person for all members of the household, and the range of possible relationships includes husband, wife, and unmarried partner (as a different category than roommate or other nonrelative). Thus, we identify individuals in same-sex couples in the following way: households with an adult who is the same sex as the primary reference person and whose relationship to the primary reference person is described as spouse or unmarried partner. We restrict our attention to individuals age 18 to 65 who were interviewed between 2009 and 2018.[7] Our final sample includes 73,000 women and 69,641 men in same-sex couples, as well as 10,809,885 men and women in different-sex couples.

### 2.2 The National Health Interview Survey (NHIS)

The main disadvantage of using ACS data is that it is not possible to identify single LGBQ individuals without a partner or same-sex couples who do not live together. Furthermore, since there is no individual-level information on sexual orientation, researchers cannot identify bisexual individuals in different-sex (or same-sex) couples. In order to address these limitations, we have analyzed data from the NHIS. The NHIS is a household, face-to-face health survey conducted by the National Center for Health Statistics of approximately 87,500 people in 35,000 households each year. The NHIS sample is designed to be representative of the U.S. civilian, non-institutionalized population. Interviewers collect information from family reference adults on the household, socio-demographic characteristics, and health indicators for all persons in the selected households. In addition, extensive information (including employment status and occupation) is collected on one randomly selected sample adult and one sample child from each family.

---

[7] We study the ACS data collected between 2009 and 2018 because information on the bachelor's degree field of study is available starting in 2009. Moreover, the U.S. Census Bureau implemented several changed between 2007 and 2008 to address concerns about misclassification errors and increase data quality (33, 34). In addition, observations with imputed sex or relationship to the primary reference person have been dropped to further reduce measurement errors (27, 33, 35).

From 2013, sample adults were asked whether they identified as straight, gay/lesbian, bisexual, or 'something else'. These data are publicly available through IPUMS-HealthSurveys at the University of Minnesota (26). Between 2013 and 2018, our final sample with information on self-reported sexual orientation and occupation includes 67,367 heterosexual women (age 18 to 65), 59,732 heterosexual men, 1,213 lesbian\gay women, 1,524 gay men, 1,113 bisexual women, and 426 bisexual men. The sample also includes 279 women and 208 men who identified with another sexual orientation category.

## 2.3 Methodology

We identify two key measures of representation in STEM fields in the ACS and NHIS: STEM degrees and STEM occupations. Information on STEM degrees is only available in the ACS; individuals were asked to identify the specific major of any bachelor's degrees each individual in the household had received. We code fields of study as being in STEM based on the individual's primary or first bachelor's degree. STEM occupations are instead observed in both our datasets.

As explained in detail in the Appendix, we follow the Department of Commerce and Bureau of Labor Statistics definitions to determine which degrees and occupations are in STEM fields. There are several reasonable alternative definitions of STEM degrees and STEM occupations. For example, some scholars include economics and finance degrees and professions in STEM. For our core definitions we do not code degrees in biology, health, economics or finance as STEM degrees. We also do not include health and medical professions or economics and finance professions in the definition of STEM occupations. Results with alternative definitions can be found in Appendix Table B1.

We start by presenting descriptive statistics and mean comparisons in Tables 1-2. We then report estimates from ordinary least squares models in Table 3. We estimate standard errors that are robust to heteroscedasticity, and we use the survey sample weights throughout. We account for the NHIS complex sample design by using the command *svy* in Stata 15 to include information on primary sampling units and strata. Details on all the variables used in the empirical analysis can be found in the Appendix.

## 3. Descriptive statistics

We begin by presenting the unconditional weighted means of our key outcome variables separately by couple type. Because of the large and well-documented gender gap in STEM, we present results separately for men and women. For women in the left columns of Table 1, the ACS data indicate that 14 percent of women in same-sex couples earned a bachelor's degree in a STEM field, while a nearly identical 13.9 percent of women in different-sex couples earned a bachelor's degree a STEM field, resulting in essentially no gap in STEM degrees between the two groups. When we examine STEM occupations, however, we observe that a larger share of women in same-sex couples are in STEM occupations compared to women in different-sex couples (5.0 versus 3.2 percent). The difference in these two means is statistically significant at the one percent level.

For men in the right columns of Table 1 we find a notably different pattern from the one for women. Specifically, we find a large gap in STEM degrees: while 22.8 percent of men in same-sex couples earned a bachelor's degree in a STEM field, the associated share for men in different-sex couples was 34.8 percent, or 12 percentage points higher – a difference that is statistically significant at the one percent level. Interestingly, the gap in STEM degrees between men in same-sex couples and men in different-sex couples is also observed for STEM occupations. Although the size of the STEM occupation gap by couple type for men is smaller (1.1 percentage points), it is statistically significant at the one percent level.[8-9]

We present the associated evidence on STEM occupations from the 2013-2018 NHIS data in Table 2. The NHIS patterns for women by self-reported sexual orientation in the left columns of Table 2 largely confirm the couples-based ACS patterns observed in Table 1. Specifically, we find that 3 percent of heterosexual women in the NHIS are in STEM occupations; the associated share for self-identified lesbian women is 3.4 percent, while 3.7 percent of self-identified bisexual women and 2.7 percent of women who describe their sexual orientation as 'something else' are observed in STEM occupations. Sample sizes in the NHIS are smaller than in the ACS, and none of these differences between the non-heterosexual groups and the heterosexual women is statistically significant.

For men in the NHIS in the right columns of Table 2 we find that 8.8 percent of self-identified heterosexual men are in STEM occupations, while only 6.5 percent of self-identified gay men are in STEM occupations, a difference in means of 2.3 percentage points that is statistically significant at the one percent level. This pattern is qualitatively identical to the ACS couples-based gap in STEM occupations documented in Table 1. We also observe that self-identified bisexual men and men who describe their sexual orientation as 'something else' are less likely to be in STEM occupations than heterosexual men, though these differences in means are not statistically significant due to small sample sizes.

---

[8] Table 1 also highlights the presence of a large gender gap in STEM degrees and STEM occupations, and it indicates that far fewer people are in STEM occupations than are observed to have STEM degrees, a fact that has been previously documented (36).

[9] Appendix Table B1 presents the associated means for STEM degrees and STEM occupations additionally disaggregated by race and age groups. Asian people are much more likely to have STEM degrees and to work in STEM occupations than white or black individuals. Notably, the gap between individuals in same-sex and different-sex couples in STEM outcomes for women are largely invariant to race. In contrast, the gap in STEM degrees and STEM occupations between Asian men in same-sex couples and Asian men in different-sex couples is much larger than the associated gaps between men of other races in same-sex couples and men of other races in different-sex couples. In addition, the gaps between individuals in same-sex and different-sex couples do not vary substantially across age groups.

## 4. Multivariate analysis

In addition to documenting the size of the unadjusted gaps in STEM degrees and occupations by sexual orientation, it is also interesting to understand the extent to which these differences can be explained by differences across groups in observable characteristics. Put differently, the STEM gaps documented in Tables 1-2 are unconditional gaps; in this section, we examine whether the differences in STEM degree and STEM occupation persist once we control for age, race, ethnicity, and location. Doing so effectively asks whether sexual minorities who are otherwise similar to heterosexual individuals have different rates of STEM representation.[10] We report the coefficient on the sexual minority variables, and in each case the relevant excluded category is the dummy variable for the majority group (individuals in different-sex couples in the ACS, self-identified heterosexual individuals in the NHIS).

Table 3 presents the multivariate regression results. The patterns largely confirm that the unadjusted gaps in STEM outcomes survive adjustment for the aforementioned observable characteristics. For example, column 1 shows that women in same-sex couples are 1.2 percentage points more likely to have STEM degrees compared to women in different-sex couples with the same age, race/ethnicity, and location, and this estimate is statistically significant at the one percent level. For men in same-sex couples compared to men in different-sex couples, the raw gap documented in Table 1 of 12 percentage points falls slightly to 10.8 percentage points once we adjust for demographic characteristics and location, though this estimate remains statistically significant at the one percent level. The patterns for STEM occupations in columns 3 and 4 are qualitatively similar: we continue to find that women in same-sex couples are slightly more represented in STEM occupations compared to women in different-sex couples, while the opposite is true for men in same-sex couples compared to men in different-sex couples.

Columns 5 and 6 of Table 4 perform the same regression adjustment exercise for self-identified sexual minorities in the NHIS. Here too we observe that the patterns observed in the unconditional differences in means are also observed in the regression estimates. Specifically, gay men are 1.9 percentage points less likely to be in STEM occupations than otherwise similar heterosexual men with the same age, race/ethnicity, and location, and this estimate is statistically significant at the five percent level. None of the other estimates on the sexual minority indicator variables is statistically significant for women or for men due to the large standard errors, thus highlighting the relatively small sample sizes in the NHIS.

---

[10] Nevertheless, it is important to emphasize that we are not accounting for different selection into higher education and employment by sexual orientation or couple type. Indeed, as shown in Appendix Table B1, individuals in same-sex couples have different levels of education, labor force participation, and employment than individuals in different-sex couples. Therefore, we are not claiming that the results in Table 3 have any causal meaning, we are only presenting estimates conditional on demographic characteristics and location, and we are not controlling for the fact that LGBQ individuals who get a bachelor's degree or enter into the labor force might be systematically different than heterosexual individuals.

Importantly, we note that the estimated gaps in the likelihood of STEM occupations between lesbian women and heterosexual women and between gay men and heterosexual men in the NHIS are similar to the regression-adjusted differences for the same outcome when we compare couples in columns 3 and 4 of Table 4. Lesbian women in the NHIS are estimated to be 0.6 percentage points more likely than similarly situated heterosexual women to have a STEM occupation, while the associated difference in column 3 for women in same-sex couples compared to women in different-sex couples is 1.9 percentage points. Gay men in the NHIS are estimated to be 1.9 percentage points less likely than similarly situated heterosexual men to have a STEM occupation, while the associated difference in column 4 for men in same-sex couples compared to men in different-sex couples indicates a 1.4 percentage point gap in the same direction. Moreover, both estimates comparing sexual minority men to heterosexual men across the ACS and NHIS are statistically significant, suggesting that there is a robust association between sexual orientation and STEM (under)representation for men in two independently drawn datasets.[11]

## 5. Relationship with Gender Gap in STEM

Extensive research has documented a robust gender gap in STEM degrees and STEM occupations: men are much more likely to be represented in STEM than women (2). Indeed, the data we analyze confirm that the gender gap in STEM is pervasive, affects both heterosexual and sexual minority women, and is larger than the associated gap between sexual minority men and heterosexual men. A natural question is whether the gap in STEM fields experienced by gay men is systematically related to the gender gap in STEM. Prior research has documented occupational sorting by gay men into female-dominated occupations (27, 28). Is this also the case in STEM?

Figure 1 presents evidence that the mechanisms underlying the gender gap in STEM may be related to those driving the gap in STEM between gay and heterosexual men. Specifically, we plot in the left panel of Figure 1 the relationship between the gender gap in STEM degrees and the gay male/heterosexual male gap in STEM degrees: the x-axis is the share of individuals in the STEM degree field that are women, and the y-axis is the share of coupled men in the STEM degree that are in same-sex couples. Each data point is a unique STEM degree field. There is a clear positive relationship between the share female in the STEM degree and the share of coupled men that is gay in the STEM degree. Moreover, the right panel of Figure 1 shows that the same relationship is observed for STEM occupations. Appendix Figure B1 shows that these positive relationships are unique to men in same-sex couples: there is no relationship (or a weakly negative one) observed when we change the y-axis to the share of coupled women in the STEM degree or STEM occupation that are women in same-sex couples.

---

[11] A large body of research confirms that gay men and lesbian women are less likely to have children present in the household than heterosexual individuals (27), and this may contribute to differences in labor force attainment or specialization in STEM fields that are not captured in our unconditional comparisons in Tables 1-2 or in our regression-adjusted comparisons in Table 3. Despite this, the results in Appendix Tables B2-B4 show that the main findings in Table 3 are not sensitive to controls for the presence of children in the household: the gaps relating to sexual orientation are qualitatively identical when we account for differences in childrearing responsibilities.

Taken together, these patterns are highly suggestive that the mechanisms underlying the very large gender gap in STEM fields are related to the factors driving the associated gap in STEM fields between gay men and heterosexual men. The patterns also suggest that policies to improve representation of women in STEM fields may have the associated benefit of increasing representation of gay men in STEM fields, and vice versa.

## 6. Conclusions

In this paper, we have provided the most comprehensive evidence on the representation of LGBQ individuals in STEM fields in the United States using large nationally representative datasets. We document the existence of a large gap in STEM degrees between men in same-sex couples and men in different-sex couples in the nationally representative American Community Surveys – a gap that is larger than the black/white STEM gap but smaller than the gender STEM gap. Moreover, we show that this gap in STEM degrees between men in same-sex couples and men in different-sex couples is observed for STEM occupations and is highly robust to controls for observable characteristics. It also replicates in independently drawn data from the National Health Interview Survey with individual level information on sexual orientation. Finally, we document that the gap in STEM degrees and occupations for men in same-sex couples is systematically related to the gender gap in STEM degrees and occupations.

We hope that this work will emphasize the importance of focusing on sexual orientation in addition to sex, race, ethnicity and disability when discussing the status of minorities in STEM fields. And while we cannot directly comment on STEM representation differences associated with transgender status due to data limitations, our work highlights the need for more large nationally representative data on both sexual minorities and gender minorities in STEM, in particular bisexual, queer and transgender individuals, to better understand their representation in undergraduate and graduate programs, in academia, and in the private sector, as well as the specific barriers and challenges faced by these groups. An important step - currently under discussion at the NSF (29) - would be to regularly include sexual orientation and gender identity measures in NSF surveys such as the Survey of Doctorate Recipients and the National Survey of College Graduates (1).

Finally, there are several areas and best practices that have been identified to foster representation of LGBTQ members in STEM fields. Researchers have already emphasized the importance of role models, representation, community, and equal treatment from employers (11, 30). Campaigns such as 500 Queer Scientists (31) and associations such as the National Organization of Gay and Lesbian Scientists and Technical Professionals are actively increasing visibility and supporting LGBTQ STEM workers. Federal agencies and universities could include LGBTQ representation into their diversity objectives (21). Furthermore, promoting gender parity could contribute to creating a welcoming and tolerant workplace climate also for sexual and gender minorities (17). More generally, fostering the use of gender-neutral pronouns could lead to more positive attitudes towards women and LGBTQ individuals (32).

# References

1. J. B. Freeman, A. P. Romero, L. Durso, "RE: National Science Foundation; Intent To Seek Approval To Renew An Information Collection For Three Years; Notice and request for comments; 2019 Survey of Doctorate Recipients" (2018).
2. NSF, "Women, Minorities, and Persons with Disabilities in Science and Engineering: 2019" (2019).
3. L. Guiso, F. Monte, P. Sapienza, L. Zingales, Culture, Gender, and Math. *Sci. Mag.* **320**, 1164–1165 (2008).
4. MAP, "Equality Maps: State Nondiscrimination Laws" (2020).
5. R. W. S. Coulter, K. S. Kenst, D. J. Bowen, Scout, Research funded by the National Institutes of Health on the health of lesbian, gay, bisexual, and transgender populations. *Am. J. Public Health* **104**, e105–e112 (2014).
6. E. A. Cech, W. R. Rothwell, LGBT Workplace Inequality in the Federal Workforce: Intersectional Processes, Organizational Contexts, and Turnover Considerations. *ILR Rev.* **73**, 25–60 (2020).
7. K. Arney, It's great if you're straight? *Sci. Careers* **August** (2004).
8. D. Bilimoria, A. J. Stewart, "Don't Ask, Don't Tell": The Academic Climate for Lesbian, Gay, Bisexual, and Transgender Faculty in Science and Engineering. *NWSA J.* **21**, 85–103 (2009).
9. E. A. Cech, T. J. Waidzunas, Navigating the heteronormativity of engineering: The experiences of lesbian, gay, and bisexual students. *Eng. Stud.* **3**, 1–24 (2011).
10. E. V. Patridge, R. S. Barthelemy, S. R. Rankin, Factors impacting the academic climate for LGBQ STEM faculty. *J. Women Minor. Sci. Eng.* **20**, 75–98 (2014).
11. APS, "LGBT Climate in Physics: Building an Inclusive Community" (2016).
12. J. G. Stout, H. M. Wright, Lesbian, Gay, Bisexual, Transgender, and Queer Students' Sense of Belonging in Computing: An Intersectional Approach. *Comput. Sci. Eng.* **18**, 24–30 (2016).
13. L. Wang, LGBT chemists seek a place at the bench. *Chem. Eng. News* **94**, 18–20 (2016).
14. E. Wagner, Employees Report Threatening Anti-LGBT Harassment, Retaliation at National Science Foundation. *Gov. Exec.* (2019).
15. M. J. Eliason, J. DeJoseph, S. Dibble, S. Deevey, P. Chinn, Lesbian, gay, bisexual, transgender, and queer/questioning nurses' experiences in the workplace. *J. Prof. Nurs.* **27**, 237–244 (2011).
16. M. J. Eliason, S. L. Dibble, P. A. Robertson, Lesbian, Gay, Bisexual, and Transgender (LGBT) Physicians' Experiences in the Workplace. *J. Homosex.* **58**, 1355–1371 (2011).
17. J. B. Yoder, A. Mattheis, Queer in STEM: Workplace Experiences Reported in a National Survey of LGBTQA Individuals in Science, Technology, Engineering, and Mathematics Careers. *J. Homosex.* **63**, 1–27 (2016).
18. A. Mattheis, D. C. R. De Arellano, J. B. Yoder, A Model of Queer STEM Identity in the Workplace. *J. Homosex.* **0**, 1–25 (2019).
19. B. E. Hughes, Coming out in STEM: Factors affecting retention of sexual minority STEM students. *Sci. Adv.* **4**, 1–6 (2018).
20. C. S. Carpenter, New evidence on gay and lesbian household incomes. *Contemp. Econ. Policy* **22**, 78–94 (2004).
21. J. Freeman, LGBTQ scientists are still left out. *Nature* **559**, 27–28 (2018).
22. A. P. Carnevale, N. Smith, M. Melton, Stem. *Cew-georget. Univ.*, 1–112 (2011).
23. Executive Office of the President, "Engage to Excel: Producing One Million Additional College Graduates with Degrees in Science, Technology, Engineering, and Mathematics." (2012).
24. S. Ruggles, *et al.*, American Community Survey Integrated Public Use Microdata Series. *IPUMS USA* **10.0** (2020).
25. D. A. Black, G. Gates, S. Sanders, L. Taylor, Demographics of the Gay and Lesbian Population in the United States : Evidence from Available Systematic Data Sources. *Demography* **37**, 139–154 (2000).
26. L. A. Blewett, J. A. R. Drew, M. L. King, K. C. W. Williams, IPUMS Health Surveys: National Health Interview Survey. *IPUMS Heal. Surv.* **6.4** (2020).
27. D. A. Black, S. G. Sanders, L. J. Taylor, The economics of lesbian and gay families. *J. Econ. Perspect.* **21**, 53–70 (2007).
28. E. Plug, D. Webbink, N. Martin, Sexual Orientation, Prejudice, and Segregation. *J. Labor Econ.* **32**, 123–159 (2014).
29. K. Langin, NSF moves to pilot LGBT questions on national workforce surveys. *Sci. Careers* **November** (2018).
30. B. Barres, B. Montague-Hellen, J. Yoder, Coming out: The experience of LGBT+ people in STEM. *Genome*

*Biol.* **18**, 62 (2017).
31. L. Esposito, S. Edgerton, 500 Queer Scientists (2019) (May 4, 2020).
32. M. Tavits, E. O. Pérez, Language influences mass opinion toward gender and LGBT equality. *Proc. Natl. Acad. Sci. U. S. A.* **116**, 16781–16786 (2019).
33. G. J. Gates, M. D. Steinberger, Same-sex Unmarried Partner Couples in the American Community Survey: The Role of Misreporting, Miscoding and Misallocation. *Annu. Meet. Popul. Assoc. Am.* **Mimeo** (2007).
34. U.S. Census, Frequently Asked Questions About Same-Sex Couple Households. *U.S. Census* **August**, 1–4 (2013).
35. T. J. DeMaio, N. Bates, M. O'Connell, Exploring Measurement Error Issues in Reporting of Same-Sex Couples. *Public Opin. Q.* **77**, 145–158 (2013).
36. U.S. Census, "Census Bureau Reports Majority of STEM College Graduates Do Not Work in STEM Occupations" (2014).

**Table 1: STEM education and occupation by type of couple, ACS.**

|  | Women | | | Men | | |
|---|---|---|---|---|---|---|
|  | In same-sex couples | In different-sex couples | Gap | In same-sex couples | In different-sex couples | Gap |
| STEM degree | 0.140 | 0.139 | 0.001 | 0.228 | 0.348 | -0.120*** |
| STEM occupation | 0.050 | 0.032 | 0.018*** | 0.084 | 0.095 | -0.011*** |
| Observations | 73,000 | 5,572,796 |  | 69,641 | 5,237,089 |  |

Notes: weighed statistics. Sample includes individuals (age 18-65) in a same-sex or different-sex couples. This table includes both the household head and the unmarried partner or married spouse in same-sex or different-sex couples. Some individuals age 18-65 may be partnered with individuals younger than 18 or older than 65, thus the sample size for men and women in different-sex couples is different. *STEM degree* only reported for respondents with a bachelor's degree. *STEM degree* does not include social sciences, finance, teacher education, and health. *STEM occupation* missing for unemployed individuals, with no work experience in the 5 years preceding the interview or earlier, or if they had never worked. Teachers and health workers not counted as STEM workers. All variables are defined in detail in the Appendix. "Observations" refers to the total number of respondents in the relevant sub-group. Source: ACS 2009-2018. * $p < 0.10$, ** $p < 0.05$, *** $p < 0.01$

**Table 2: STEM occupation by sexual orientation, NHIS.**

|  | Women | | | Men | | |
|---|---|---|---|---|---|---|
|  | N | STEM occupation | Gap with straight women | N | STEM occupation | Gap with straight men |
| Straight | 67,367 | 0.030 |  | 59,732 | 0.088 |  |
| Lesbian or gay | 1,213 | 0.034 | 0.004 | 1,524 | 0.065 | -0.023*** |
| Bisexual | 1,113 | 0.037 | 0.007 | 426 | 0.077 | -0.011 |
| Something else | 279 | 0.027 | -0.003 | 208 | 0.067 | -0.021 |

Notes: weighed statistics. Sample includes all sample adults (age 18-65) who were working in the week preceding the interview, with a job or business but not at work, or who had ever worked. Respondents not in the universe, who refused to answer the occupation question or with missing information have been excluded. Teachers and health workers not counted as STEM workers. All variables are defined in detail in the Appendix. Source: NHIS 2013-2018. * $p < 0.10$, ** $p < 0.05$, *** $p < 0.01$

**Table 3: STEM education and occupation, regression adjusted estimates.**

|  | ACS | | | | NHIS | |
|---|---|---|---|---|---|---|
|  | STEM degree | | STEM occupation | | STEM occupation | |
|  | Women | Men | Women | Men | Women | Men |
|  | (1) | (2) | (3) | (4) | (5) | (6) |
| In a same-sex couple | 0.012*** | -0.108*** | 0.019*** | -0.014*** | | |
|  | (0.002) | (0.003) | (0.001) | (0.001) | | |
| Gay or lesbian |  |  |  |  | 0.006 | -0.019** |
|  |  |  |  |  | (0.006) | (0.008) |
| Bisexual |  |  |  |  | 0.008 | -0.017 |
|  |  |  |  |  | (0.007) | (0.015) |
| Something else |  |  |  |  | -0.001 | -0.018 |
|  |  |  |  |  | (0.014) | (0.020) |
| Dependent variable mean | 0.139 | 0.345 | 0.032 | 0.095 | 0.030 | 0.087 |
| R-squared | 0.030 | 0.039 | 0.015 | 0.029 | 0.012 | 0.028 |
| Observations | 2,063,090 | 1,850,340 | 4,664,190 | 4,992,047 | 69,972 | 61,890 |

Notes: The dependent variable in columns 1-2 is whether an individual received a bachelor's degree in a STEM field. The dependent variable in columns 3-6 is whether an individual used to work in a STEM occupation. See also notes in Tables 1-2. All regressions include controls for demographic characteristics (age, race, ethnicity) and location (state fixed effects in the ACS, region fixed effects in the NHIS since we do not observe state of residence in the NHIS public-use data). All variables are defined in detail in the Appendix. Weighted regressions using person weights. Standard errors in parentheses. Source: ACS 2009-2018 (Columns 1-4), and NHIS 2013-2018 (Columns 5-6). * $p < 0.10$, ** $p < 0.05$, *** $p < 0.01$

**Figure 1: Share of women and individuals in same-sex couples in STEM, ACS.**

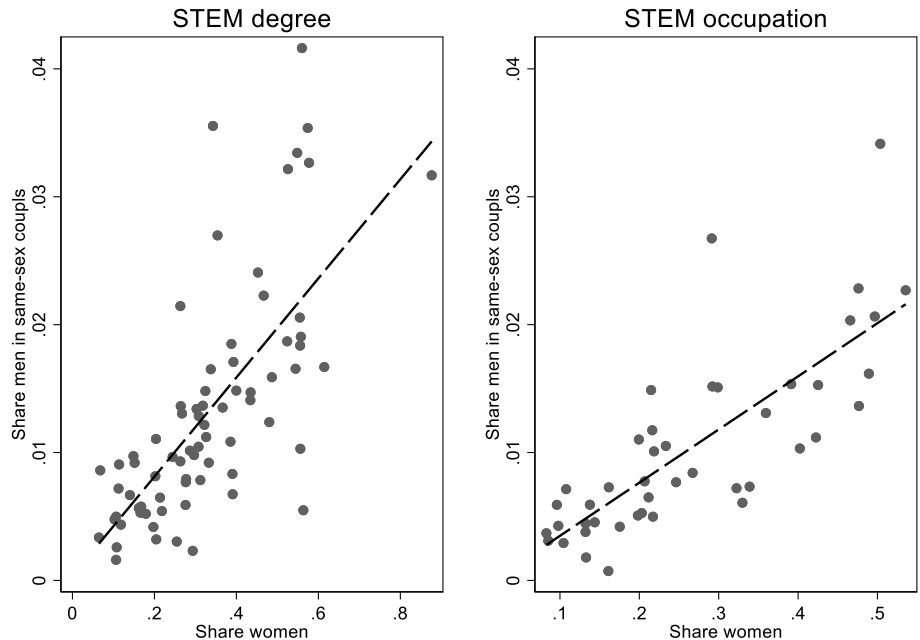

Notes: weighed statistics. See also notes in Tables 1. The vertical axis measures the share of men in same-sex couples over all coupled men in each field/occupation. The horizontal axis measures the share of women (of any marital status and relation to the household head) over all individuals in each field/occupation. Only STEM fields/occupations reported. The dashed line plots the linear fit. Source: ACS 2009-2018.

**Appendix A: Variable description**

**A.1 ACS variables**

*Sex* reports whether an individual is male or female.

*In a same-sex couple*. The ACS does not directly ask individuals about their sexual orientation. However, the ACS identifies a primary reference person, defined as "the person living or staying here in whose name this house or apartment is owned, being bought, or rented". The ACS also collects information on the relationship to the primary reference person for all members of the household, and the range of possible relationships includes husband, wife, and unmarried partner (as a different category than roommate or other nonrelative). By combining such information, we have created an indicator variable equal to one if an individual was in a same-sex couple; zero otherwise. We have coded as individuals in a same-sex couple both married individuals and individuals living with an unmarried partner.

*Higher Education* is an indicator equal to one if an individual's highest degree completed was a bachelor's degree or higher (Master's degree, Professional degree beyond a bachelor's degree, Doctoral degree); zero otherwise.

*STEM degree* is an indicator equal to one if an individual received a bachelor's degree in a STEM field; zero if they received their degree in a non-STEM field. This indicator refers to the primary field in which an individual received a bachelor's degree. This indicator has been set to missing if an individual did not receive a bachelor's degree. In line with the classification used by the U.S. Department of Commerce[12] and the U.S. Bureau of Labor Statistics,[13] the following fields have been coded as STEM:

- Agricultural Sciences
    - Animal Sciences
    - Food Science
    - Plant Science and Agronomy
    - Soil Science
- Environmental Science
- Architecture
- Communication Technologies
- Computer and Information Systems
    - Computer Programming and Data Processing
    - Computer Science
    - Information Sciences
    - Computer Information Management and Security
    - Computer Networking and Telecommunications

---

[12] Source: https://files.eric.ed.gov/fulltext/ED594354.pdf (Appendix Table 2).
[13] Source: https://www.bls.gov/oes/topics.htm#stem

- General Engineering
  - Aerospace Engineering
  - Biological Engineering
  - Architectural Engineering
  - Biomedical Engineering
  - Chemical Engineering
  - Civil Engineering
  - Computer Engineering
  - Electrical Engineering
  - Engineering Mechanics, Physics, and Science
  - Environmental Engineering
  - Geological and Geophysical Engineering
  - Industrial and Manufacturing Engineering
  - Materials Engineering and Materials Science
  - Mechanical Engineering
  - Metallurgical Engineering
  - Mining and Mineral Engineering
  - Naval Architecture and Marine Engineering
  - Nuclear Engineering
  - Petroleum Engineering
  - Miscellaneous Engineering
- Engineering Technologies
  - Engineering and Industrial Management
  - Electrical Engineering Technology
  - Industrial Production Technologies
  - Mechanical Engineering Related Technologies
  - Miscellaneous Engineering Technologies
- Biology
  - Biochemical Sciences
  - Botany
  - Molecular Biology
  - Ecology
  - Genetics
  - Microbiology
  - Pharmacology
  - Physiology
  - Zoology
  - Neuroscience
  - Miscellaneous Biology
- Mathematics

- o   Applied Mathematics
- o   Statistics and Decision Science
- Military Technologies
- Interdisciplinary and Multi-Disciplinary Studies
    - o   Nutrition Sciences
    - o   Neuroscience
    - o   Mathematics and Computer Science
    - o   Cognitive Science and Biopsychology
- Physical Sciences
    - o   Astronomy and Astrophysics
    - o   Atmospheric Sciences and Meteorology
    - o   Chemistry
    - o   Geology and Earth Science
    - o   Geosciences
    - o   Oceanography
    - o   Physics
    - o   Materials Science
- Nuclear, Industrial Radiology, and Biological Technologies
- Transportation Sciences and Technologies
- Actuarial Science
- Operations, Logistics and E-Commerce
- Management Information Systems and Statistics

*In the labor force* is an indicator equal to one if an individual was a part of the labor force, either working or seeking work, in the week preceding the interview; zero otherwise.

*Unemployed* is an indicator equal to one if an individual did not have a job, was looking for a job, and had not yet found one at the time of the interview, rather than being employed. Individuals not in the labor force have been coded as missing. Individuals who had never worked but were actively seeking their first job are considered unemployed.

*STEM occupation* is an indicator equal to one if an individual's primary occupation was in a STEM field; zero otherwise. This indicator has been set to missing if an individual was unemployed, with no work experience in the 5 years preceding the interview or earlier, or if they had never worked. Because of data limitations (no detailed codes for teachers), STEM postsecondary educators cannot be coded as STEM workers. In line with the classification used by the U.S. Department of Commerce[14] and the U.S. Bureau of Labor Statistics,[15] the following occupations have been coded as STEM:

- Management, Business, Science, and Arts Occupations

---

[14] Source: https://files.eric.ed.gov/fulltext/ED594354.pdf (Appendix Table 1).
[15] Source: https://www.bls.gov/oes/topics.htm#stem

- o Computer and information systems managers
- o Architectural and Engineering Managers
- o Natural Sciences Managers
- Computer and Mathematical Occupations
  - o Computer and Information Research Scientists
  - o Computer Systems Analysts
  - o Information security analysts
  - o Computer Programmers
  - o Software developers
  - o Software quality assurance analysts and testers
  - o Web Developers
  - o Web and digital interface designers
  - o Computer support specialists
  - o Database Administrators
  - o Network and Computer Systems Administrators
  - o Computer Network Architects
  - o Computer occupations, all other
  - o Actuaries
  - o Operations Research Analysts
  - o Other mathematical science occupations
- Architecture and Engineering Occupations
  - o Architects, Except landscape and Naval
  - o Landscape architects
  - o Surveyors, Cartographers, and Photogrammetrists
  - o Aerospace Engineers
  - o Biomedical and agricultural engineers
  - o Chemical Engineers
  - o Civil Engineers
  - o Computer Hardware Engineers
  - o Electrical and Electronics Engineers
  - o Environmental Engineers
  - o Industrial Engineers, including Health and Safety
  - o Marine Engineers and Naval Architects
  - o Materials Engineers
  - o Mechanical Engineers
  - o Petroleum, mining and geological engineers, including mining safety engineers
  - o Miscellaneous engineeers including nuclear engineers
  - o Drafters
  - o Engineering Technicians, Except Drafters
  - o Surveying and Mapping Technicians

- Life, Physical, and Social Science Occupations
    - Agricultural and Food Scientists
    - Biological Scientists
    - Conservation Scientists and Foresters
    - Other life scientists
    - Astronomers and Physicists
    - Atmospheric and Space Scientists
    - Chemists and Materials Scientists
    - Environmental scientists and specialists, including health
    - Geoscientists and hydrologists, except geographers
    - Physical Scientists, All Other
    - Agricultural and Food Science Technicians
    - Biological Technicians
    - Chemical Technicians
    - Environmental science and geoscience technicians, and nuclear technicians
    - Other life, physical, and social science technicians
    - Occupational health and safety specialists and technicians
- Sales Engineers

*Age* reports an individual's age in years at the time of the interview.

*Race*. A series of indicator variables has been constructed to record an individual's race: White, Black, Asian, or other races. Asian includes Chinese, Japanese, Other Asian or Pacific Islander. Other races include American Indian, Alaska Native, other race not listed, or individuals who selected two or three major races.

*Hispanic* is an indicator equal to one if an individual self-identified as Mexican, Puerto Rican, Cuban, or Other Hispanic; zero otherwise.

*Fertility* is an indicator equal to one if an individual had one or more own children (of any age or marital status) living in the household at the time of the interview, zero otherwise. This indicator includes step-children and adopted children as well as biological children.

**A.2 NHIS variables**

*Sex* reports whether an individual was male or female. The original NHIS variable is available for all individuals.

*Sexual orientation* reports an individual's sexual orientation. The original NHIS variable is available for sample adults age 18 or older. The original question is the following:

Which of the following best represents how you think of yourself?

- Gay [lesbian or gay when asked to women]
- Straight, that is, not gay

- Bisexual
- Something else
- I don't know the answer
- Refused

*Higher Education* is an indicator equal to one if an individual's highest degree completed was a bachelor's degree or higher (Master's degree, Professional degree beyond a bachelor's degree, Doctoral degree); zero otherwise. This indicator has been set to missing if an individual refused to answer, did not know, or if the original NHIS variable is missing. The original NHIS variable is available for all individuals age 5 or older.

*In the labor force* is an indicator equal to one if an individual was a part of the labor force, either working (Working for pay at job/business; Working, without pay, at job/business; With job, but not at work) or seeking work (unemployed) in the week preceding the interview; zero otherwise. This indicator has been set to missing if an individual refused to answer, did not know, or if the original NHIS variable is missing. The original NHIS variable is available for all individuals age 18 or older.

*Unemployed* is an indicator equal to one if an individual did not have a job, was looking for a job, and had not yet found one at the time of the interview, rather than being employed. Individuals not in the labor force have been coded as missing. This indicator has been set to missing if an individual refused to answer, did not know, or if the original NHIS variable is missing. The original NHIS variable is available for all individuals age 18 or older.

*STEM occupation* is an indicator equal to one if an individual's primary occupation was in a STEM field; zero otherwise. This indicator has been set to missing for individuals not in the universe, if an individual refused to answer, did not know, or if the original NHIS variable is missing. The original NHIS variable is available for sample adults age 18 or older who were working at a paying job in the week preceding the interview; with a job or business but not at work; working at a non-paying job in the week preceding the interview; or who had ever worked. The following occupations have been coded as STEM:

- Computer and Mathematical Occupations
    - Computer specialists
    - Mathematical science occupations
- Architecture and Engineering Occupations
    - Architects, surveyors, and cartographers
    - Engineers
    - Drafters, engineering, and mapping technicians
- Life, Physical, and Social Science Occupations
    - Life scientists
    - Physical scientists

*Age* reports an individual's age in years at the time of the interview (top coded for 85 years or older). The original NHIS variable is available for all individuals.

*Race*. A series of indicator variables has been constructed to record an individual's race: White, Black, Asian, or other races. Asian includes Chinese, Filipino, Asian Indian, or Other Asian. Other races include American Indian or Alaska Native, multiple races with no primary race selected, or individuals whose primary race was not releasable. The original NHIS variable is available for all individuals.

*Hispanic* is an indicator equal to one if an individual self-identified as Hispanic or Latino (Puerto Rican, Cuban or Cuban American, Dominican, Mexican or Mexican American, Central or South American, Other Latin American, Other Hispanic or Latino); zero otherwise. The original NHIS variable is available for all individuals.

*Fertility* is an indicator equal to one if an individual had one or more own children (of any age or marital status) living in the household at the time of the interview, zero otherwise. This indicator includes step-children and adopted children as well as biological children. The original NHIS variable is available for all individuals.

# Appendix B: Additional figures and tables

## Table B1: STEM education and occupation by type of couple. ACS extensions.

### Panel A: all individuals in couples.

|  | Women | | | Men | | |
|---|---|---|---|---|---|---|
|  | Same-sex | Different-sex | Gap | Same-sex | Different-sex | Gap |
| *STEM degree:* | | | | | | |
| As in the main table | 0.140 | 0.139 | 0.001 | 0.228 | 0.348 | -0.120*** |
| With STEM education | 0.144 | 0.143 | 0.001 | 0.230 | 0.351 | -0.121*** |
| Without biology | 0.091 | 0.093 | -0.002 | 0.172 | 0.297 | -0.125*** |
| Without food/nutrition | 0.137 | 0.135 | 0.002 | 0.226 | 0.347 | -0.121*** |
| With secondary field | 0.149 | 0.145 | 0.004 | 0.238 | 0.357 | -0.119*** |
| *Additional variables:* | | | | | | |
| Bachelor's degree or more | 0.448 | 0.359 | 0.089*** | 0.487 | 0.339 | 0.148*** |
| In the labor force | 0.842 | 0.706 | 0.136*** | 0.852 | 0.884 | -0.032*** |
| Unemployed | 0.048 | 0.050 | -0.002** | 0.049 | 0.044 | 0.005*** |
| Observations | 73,000 | 5,572,796 |  | 69,641 | 5,237,089 |  |

### Panel B: by race.

|  | In same-sex couples | | | In different-sex couples | | |
|---|---|---|---|---|---|---|
|  | White | Black | Asian | White | Black | Asian |
| *Women:* | | | | | | |
| STEM degree | 0.136 | 0.125 | 0.289 | 0.120 | 0.115 | 0.301 |
| STEM occupation | 0.050 | 0.030 | 0.122 | 0.028 | 0.024 | 0.106 |
| Observations | 60,801 | 5,384 | 2,018 | 4,574,437 | 331,965 | 340,241 |
| *Men:* | | | | | | |
| STEM degree | 0.216 | 0.206 | 0.388 | 0.321 | 0.275 | 0.613 |
| STEM occupation | 0.087 | 0.043 | 0.152 | 0.091 | 0.055 | 0.255 |
| Observations | 58,563 | 3,546 | 3,055 | 4,308,142 | 348,464 | 270,806 |

### Panel B: by age groups.

|  | In same-sex couples | | | In different-sex couples | | |
|---|---|---|---|---|---|---|
|  | 18-34 | 35-49 | 50-65 | 18-34 | 35-49 | 50-65 |
| *Women:* | | | | | | |
| STEM degree | 0.147 | 0.144 | 0.146 | 0.163 | 0.146 | 0.152 |
| STEM occupation | 0.042 | 0.055 | 0.049 | 0.034 | 0.035 | 0.035 |
| Observations | 19,659 | 25,864 | 47,559 | 1,204,997 | 1,996,685 | 3,358,213 |
| *Men:* | | | | | | |
| STEM degree | 0.251 | 0.230 | 0.235 | 0.350 | 0.358 | 0.356 |
| STEM occupation | 0.080 | 0.092 | 0.087 | 0.096 | 0.103 | 0.100 |
| Observations | 15,528 | 26,533 | 44,309 | 966,979 | 1,902,722 | 3,019,751 |

Notes: weighed statistics. See also notes in Tables 1. *STEM degree with education* counts as STEM also: Mathematics Teacher Education; Science and Computer Teacher Education. *STEM degree without biology* does not count as STEM: Biology; Biochemical Sciences; Botany; Molecular Biology; Ecology; Genetics; Microbiology; Pharmacology; Physiology; Zoology; Neuroscience; Miscellaneous Biology. *STEM degree without food/nutrition* does not count as STEM: Food Science; Nutrition Sciences. *STEM degree with secondary field* indicates whether an individual received a bachelor's degree with a primary or secondary concentration in STEM. Unemployment rate is computed only among the individuals in the labor force. "Observations" refers to the total number of respondents in the relevant sub-group. Source: ACS 2009-2018. * $p < 0.10$, ** $p < 0.05$, *** $p < 0.01$

**Table B2: STEM degree by type of couple. ACS additional regressions.**

|  | Women | | | | Men | | | |
| --- | --- | --- | --- | --- | --- | --- | --- | --- |
|  | (1) | (2) | (3) | (4) | (5) | (6) | (7) | (8) |
| Same-sex couple | 0.001 | 0.014*** | 0.012*** | 0.010*** | -0.120*** | -0.112*** | -0.108*** | -0.106*** |
|  | (0.002) | (0.002) | (0.002) | (0.002) | (0.003) | (0.003) | (0.003) | (0.003) |
| Demographic | No | Yes | Yes | Yes | No | Yes | Yes | Yes |
| State FE | No | No | Yes | Yes | No | No | Yes | Yes |
| Fertility | No | No | No | Yes | No | No | No | Yes |
| Mean dep var | 0.139 | 0.139 | 0.139 | 0.139 | 0.345 | 0.345 | 0.345 | 0.345 |
| R-squared | 0.000 | 0.028 | 0.030 | 0.030 | 0.001 | 0.035 | 0.039 | 0.039 |
| Observations | 2,063,090 | 2,063,090 | 2,063,090 | 2,063,090 | 1,850,340 | 1,850,340 | 1,850,340 | 1,850,340 |

Notes: The dependent variable is whether an individual received a bachelor's degree in a STEM field. See also notes in Table 1. Demographic characteristics: age, race, ethnicity. Fertility is an indicator equal to one if an individual had a child living in their household. All variables are defined in detail in the Online Appendix. Weighted regressions using person weights. Robust standard errors in parentheses. Source: ACS 2009-2018. * $p < 0.10$, ** $p < 0.05$, *** $p < 0.01$

**Table B3: STEM occupation by type of couple. ACS additional regressions.**

|  | Women | | | | Men | | | |
| --- | --- | --- | --- | --- | --- | --- | --- | --- |
|  | (1) | (2) | (3) | (4) | (5) | (6) | (7) | (8) |
| Same-sex couple | 0.018*** | 0.020*** | 0.019*** | 0.017*** | -0.010*** | -0.011*** | -0.014*** | -0.016*** |
|  | (0.001) | (0.001) | (0.001) | (0.001) | (0.001) | (0.001) | (0.001) | (0.001) |
| Demographic | No | Yes | Yes | Yes | No | Yes | Yes | Yes |
| State FE | No | No | Yes | Yes | No | No | Yes | Yes |
| Fertility | No | No | No | Yes | No | No | No | Yes |
| Mean dep var | 0.032 | 0.032 | 0.032 | 0.032 | 0.095 | 0.095 | 0.095 | 0.095 |
| R-squared | 0.000 | 0.013 | 0.015 | 0.015 | 0.000 | 0.024 | 0.029 | 0.029 |
| Observations | 4,664,190 | 4,664,190 | 4,664,190 | 4,664,190 | 4,992,047 | 4,992,047 | 4,992,047 | 4,992,047 |

Notes: The dependent variable is whether an individual used to work in a STEM occupation. See also notes in Table 1. Demographic characteristics: age, race, ethnicity. Fertility is an indicator equal to one if an individual had an own child living in their household. All variables are defined in detail in the Online Appendix. Weighted regressions using person weights. Robust standard errors in parentheses. Source: ACS 2009-2018. * $p < 0.10$, ** $p < 0.05$, *** $p < 0.01$

**Table B4: STEM occupation by sexual orientation. NHIS additional regressions.**

|  | Women | | | | Men | | | |
|---|---|---|---|---|---|---|---|---|
|  | (1) | (2) | (3) | (4) | (5) | (6) | (7) | (8) |
| Gay or lesbian | 0.004 | 0.006 | 0.006 | 0.003 | -0.023*** | -0.018** | -0.019** | -0.014* |
|  | (0.006) | (0.006) | (0.006) | (0.006) | (0.008) | (0.008) | (0.008) | (0.008) |
| Bisexual | 0.007 | 0.008 | 0.008 | 0.007 | -0.011 | -0.017 | -0.017 | -0.014 |
|  | (0.007) | (0.007) | (0.007) | (0.007) | (0.016) | (0.015) | (0.015) | (0.015) |
| Something else | -0.003 | -0.001 | -0.001 | -0.004 | -0.021 | -0.018 | -0.018 | -0.015 |
|  | (0.015) | (0.015) | (0.014) | (0.014) | (0.019) | (0.020) | (0.020) | (0.020) |
| Demographic | No | Yes | Yes | Yes | No | Yes | Yes | Yes |
| Region FE | No | No | Yes | Yes | No | No | Yes | Yes |
| Fertility | No | No | No | Yes | No | No | No | Yes |
| Mean dep var | 0.030 | 0.030 | 0.030 | 0.030 | 0.087 | 0.087 | 0.087 | 0.087 |
| R-squared | 0.000 | 0.012 | 0.012 | 0.013 | 0.000 | 0.028 | 0.028 | 0.029 |
| Observations | 69,972 | 69,972 | 69,972 | 69,972 | 61,890 | 61,890 | 61,890 | 61,890 |

Notes: The dependent variable is whether an individual used to work in a STEM occupation. See also notes in Table 2. Demographic characteristics: age, race, ethnicity. Fertility is an indicator equal to one if an individual had an own child living in their household. All variables are defined in detail in the Online Appendix. Weighted regressions using person weights. Standard errors in parentheses computed using the Stata command *svy* to account for the NHIS sampling design. Source: ACS 2009-2018. * $p < 0.10$, ** $p < 0.05$, *** $p < 0.01$

**Figure B1: Share of women and individuals in same-sex couples in STEM education, ACS.**

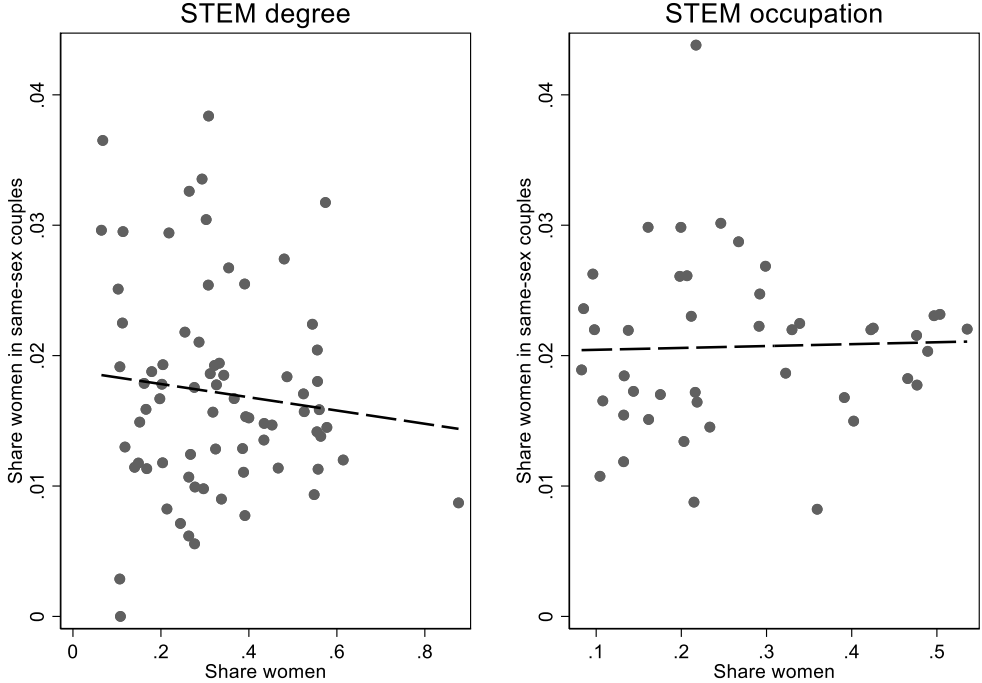

Notes: weighed statistics. See also notes in Tables 1. The vertical axis measures the share of women in same-sex couples over all coupled women in same-sex or different-sex couples in each field/occupation. The horizontal axis measures the share of women (of any marital status and relation to the household head) over all individuals in each field/occupation. Only STEM fields/occupations reported. The dashed line plots the linear fit. Source: ACS 2009-2018.